\documentclass[pre,superscriptaddress,floatfix,twocolumn]{revtex4-1}
\pdfoutput=1
\usepackage{amsmath,amssymb,eucal,graphicx,color}

\usepackage{subfigure}
\usepackage{setspace}

\usepackage{textcomp}
\usepackage{booktabs,tabularx,dcolumn}
\newcolumntype{d}[1]{D{.}{.}{#1}}

\usepackage{url,hyperref}
\usepackage[usenames,dvipsnames]{xcolor}

\usepackage{placeins}

\begin{document}

\title{Current-mediated synchronization  of a pair of beating non-identical flagella}

\author{V. S. Dotsenko}
\affiliation{Sorbonne Universit\'e, CNRS, Laboratoire de Physique Th\'eorique de la Mati\`{e}re Condens\'ee, LPTMC (UMR CNRS 7600),
75252 Paris Cedex 05, France}
\affiliation{L.D.\ Landau Institute for Theoretical Physics,
           119334 Moscow, Russia}
\author{A. Maciolek}
\affiliation{Max-Planck-Institut f{\"u}r Intelligente Systeme,
  Heisenbergstra\ss e~3, D-70569 Stuttgart, Germany}
\affiliation{IV. Institut f\"ur Theoretische und Angewandte Physik,
  Universit\"at Stuttgart, D-70569 Stuttgart, Germany}
\affiliation{Institute of Physical Chemistry, Polish Academy of Sciences,
  Department III, Kasprzaka 44/52, PL-01-224 Warsaw, Poland}
  \author{G. Oshanin$^{*}$}
\affiliation{Sorbonne Universit\'e, CNRS, Laboratoire de Physique Th\'eorique de la Mati\`{e}re Condens\'ee, LPTMC (UMR CNRS 7600),
75252 Paris Cedex 05, France}
\email{oshanin@lptmc.jussieu.fr}
\author{O. Vasilyev}
\affiliation{Max-Planck-Institut f{\"u}r Intelligente Systeme,
  Heisenbergstra\ss e~3, D-70569 Stuttgart, Germany}
\affiliation{IV. Institut f\"ur Theoretische und Angewandte Physik,
  Universit\"at Stuttgart, D-70569 Stuttgart, Germany}
\author{S. Dietrich}
\affiliation{Max-Planck-Institut f{\"u}r Intelligente Systeme,
  Heisenbergstra\ss e~3, D-70569 Stuttgart, Germany}
\affiliation{IV. Institut f\"ur Theoretische und Angewandte Physik,
  Universit\"at Stuttgart, D-70569 Stuttgart, Germany}

\date{\today}

\begin{abstract}
The basic phenomenology of experimentally observed synchronization (i.e., a stochastic phase locking)
of identical, beating flagella of a biflagellate alga is known to be captured well
by a minimal model describing the dynamics of coupled, limit-cycle, noisy oscillators
(known as the noisy Kuramoto model). As demonstrated experimentally,
the amplitudes of the noise terms therein, which stem from fluctuations of the rotary motors, depend on the flagella length. Here we address the  conceptually
important question which kind of synchrony occurs
if the two flagella have different lengths such that
the noises acting on each of them have different amplitudes.
On the basis of a minimal model, too, we show that  
 a different kind of
synchrony emerges, and
 here it is
mediated by a current carrying,  steady-state; it manifests itself via correlated "drifts" of phases.
We quantify such a synchronization mechanism in terms of appropriate order parameters $Q$ and $Q_{\cal S}$ - for an ensemble of trajectories and
for a single realization of noises  of duration ${\cal S}$, respectively.
Via numerical simulations we show that both approaches become identical for long observation times ${\cal S}$.  This 
reveals an ergodic behavior and
implies
that a single-realization order parameter $Q_{\cal S}$ is suitable for experimental analysis for which ensemble averaging is not always possible.
\end{abstract}


\maketitle


\section{Introduction}

There is  experimental evidence that two beating flagella, extending from one end of the biflagellate
alga \textit{Chlamydomonas} reinhardtii,
synchronize their dynamics.
Analyzing the oscillatory intensity signals
$x_{1,2}(t) = \Gamma_{1,2}(t) \sin\left(2 \pi \theta_{1,2}(t)\right)$
(where $\Gamma_{1,2}(t)$  and $\theta_{1,2}(t)$
are the amplitudes and  the instantaneous phases of the periodic motion
of  flagella $1$ and $2$, respectively), which are 
obtained by local sampling of the video light intensity near the two flagella, Polin et al. \cite{1} observed that the
phase difference $\triangle_t = \theta_1(t) - \theta_2(t)$ contains periods of synchrony (i.e., the so-called phase locking behavior with $\Delta_t \approx {\rm const}$ \cite{11a,m2,11,11b}),
interrupted by sudden drifts of either sign. 
Referring to earlier ideas, that the hydrodynamic interactions between
eukaryotic flagella or cilia may underlie
their synchronization \cite{2,3,4,5,6,7,8,9}, Goldstein et al. \cite{10}
proposed a  phenomenological, minimal, stochastic model in which the motion of a
flagella pair is  described by two noisy phase oscillators
which move on circular trajectories and are
coupled via an antisymmetric function of  the phase difference $\triangle_t$.
In terms of the notations used in Ref. \cite{10}, the equations of motion read
\begin{align}
\label{phases}
\dot{\theta}_{1}(t) &= \nu_{1} + \pi \varepsilon \sin\left[2 \pi \left(\theta_{2}(t)  - \theta_{1}(t)\right)\right] + \zeta_{1}(t) \,, \nonumber\\
\dot{\theta}_{2}(t) &= \nu_{2} + \pi \varepsilon \sin\left[2 \pi \left(\theta_{1}(t)  - \theta_{2}(t)\right)\right] + \zeta_{2}(t) \,.
\end{align}
In Eq. \eqref{phases}, the dimensionless functions
$\theta_{1,2}(t) \in (-1/2,1/2)$ are the two phases mentioned above,
the dot denotes the time derivative,
$\nu_{1}$ and $\nu_{2}$ are the natural  frequencies (${\rm Hz}$) of the flagella $1$ and $2$, respectively, and
$\varepsilon$ (${\rm Hz}$)  is the amplitude of the coupling between the  flagella. 
This  phenomenological parameter accounts
 for the fact that fluid flows driven by beating flagella provide a hydrodynamic coupling between the latter.
Lastly,  $\zeta_{1,2}(t)$ are delta-correlated,
Gaussian white noises with zero mean and identical covariances \cite{remark,kalm}:
$\overline{\zeta_i(t) \zeta_j(t')} = 2 T_{\rm eff} \delta_{i,j} \delta(t-t')$,
where the bar denotes the average over realizations of the noises; 
$i,j \in {1,2}$; $\delta_{i,j}$ is the Kronecker symbol  while $T_{\rm eff}$  (${\rm Hz}$) 
can be  considered as an \textit{effective} "temperature",
because it defines the amplitude of the noise terms.  Importantly,  the major contribution to $T_{\rm eff}$
stems from the fluctuations of  the rotary motors of flagella  \cite{10,10a,10b}. 
Indeed, these \textit{active} fluctuations are several orders of magnitude 
larger than the thermal noise \cite{10,10b}, so that the flagella 
are not in thermal equilibrium with its bath.

We note that the sine terms in Eq. \eqref{phases}, which  link
the  time evolutions of  $\theta_1(t)$ and $\theta_2(t)$, describe the actual coupling
due to hydrodynamic interactions between the two flagella \cite{2,3,4,5,6,7,8,9} only in an  effective way.
However, up to now no other good and justified alternative to such
a phenomenological description has emerged. In particular,
explicit results presented in, e.g., Ref. \cite{9}, are based on the assumption that the distance between the two flagella is much larger than their length, which is not the case for the system studied in Ref. \cite{10}. We also note that in more complex situations of unicellular algal species bearing \textit{multiple} flagella, both the effective coupling introduced in Ref. \cite{10} and the results in Refs. \cite{2,3,4,5,6,7,8,9}
may turn out to be insufficient to describe properly all  facets of the synchronized behavior: a different synchronization scenario may be realized which is provided, e.g., by contractile fibers of the basal apparatus  \cite{9a,21}.

Solving Eq. \eqref{phases} numerically for given realizations of the noise terms, Goldstein et al. \cite{10} have found consistency between  $\Delta_t$ evolving according to Eq. \eqref{phases} and
the experimentally observed behavior of the flagella  of the biflagellate
alga \cite{1}, i.e.,  the  calculated trajectories   of
the phase difference  exhibit
essentially the same noisy synchronization interrupted by occasional phase slips.

The comparison with  experimental data has facilitated to
identify
the physically relevant
values of  the parameters entering the effective Langevin equations. In particular,
for flagella of length $\textit{l} \simeq 12 \, \mu {\rm m}$,
observations based on the dynamics of $21$ individuals and the comparison with the time series for $\Delta_t$,
spanning over an interval of $10^2$ seconds (i.e.,  containing  several thousands of beats),
have shown that $\varepsilon$ lies within the range $0.14 - 0.7$ ${\rm Hz}$ and that the effective temperature $T_{\rm eff}$ is within the range $0.05 - 0.28$ ${\rm Hz}$,
while  $\nu = (\nu_1 + \nu_2)/2 \simeq 47$ ${\rm Hz}$ and
 $\delta \nu/\nu = |\nu_1 - \nu_2|/\nu \simeq 0.004$. The experimentally obtained
values of $\varepsilon$ appear to be in line
with the theoretical prediction in Ref. \cite{9}. Importantly, the results of Ref. \cite{10}
have emphasized for the first time the essential role played by the biochemical noise
 in the dynamics of eukaryotic flagella as manifested by its  realization-to-realization fluctuations.

A more sophisticated experimental analysis has been performed in
Ref. \cite{c}, which is focused on the dependence of the coupling parameter
$\varepsilon$ and of the effective temperature $T_{\rm eff}$ on the  length \textit{l}  of the flagella. 
This enhanced experimental analysis took advantage of the ability of the
\textit{Chlamydomonas} reinhardtii
alga to shed its flagella and  to regrow  them after a
deflagellation has occurred \cite{non-equal1}. 
The flagella of length $l \simeq 10.82$ $\mu$m have been first clipped by a micropipette \cite{c}
and then left to regrow. Within $90$ minutes
the flagella reached the
length $l \simeq 11.48$ $\mu$m, which, surprisingly, exceeded the original one. The dynamics of
the slowly growing flagella has been recorded every ten minutes within time  intervals two minutes long
(within which the length of the flagella did not appreciably change).

From $19$  such experiments
it was inferred \cite{c} that, for progressively longer
 flagella, the periods of the synchronous beating of both flagella become more pronounced. Analyzing the data,
 it was shown that, as the flagella grow, the beating frequency $\nu$ decreases   $\propto \; 1/l$,  implying that the beating mechanism operates at constant power output  per  length (see Ref.  \cite{c}). The coupling parameter
$\varepsilon$ turned out  to be linearly proportional to \textit{l}, which explains the trend for a progressive increase of the synchronization periods. Therefore, the proportionality $\varepsilon \;  \propto \; l$ is in  agreement with the elastohydrodynamic scaling $\varepsilon \; \propto \; \nu^2 l^3$ as predicted in Ref. \cite{9}. Lastly, a variation of $T_{\rm eff}$ with \textit{l} has been observed.  In particular, for $l \simeq 6$ $\mu$m, $T_{\rm eff}$ was found to lie within the range $0.06 - 0.09$ {\rm Hz}, while for $l \simeq 8$ $\mu$m it lies within the range $0.04 - 0.06$ ${\rm Hz}$, which is not overlapping with the previous interval. For larger values of $l$, $T_{\rm eff}$ was shown to saturate at  a constant value of the order of $0.04$ $\rm Hz$. Therefore, $T_{\rm eff}$  evidently depends on \textit{l}, at least for sufficiently short flagella, such that it is larger for shorter flagella.
In view of the active nature of the fluctuations of
 the rotary motors, this is in line with the intuitively expected behavior \cite{22}.

Once noise appears to be a physically relevant parameter, it is natural  to explore
a wider range of  possible effects. In this sense the conceptually important question arises
 what kind of  synchronization, if any, may take place in situations in which the length of the two flagella differ.
 Such a situation may apparently be realized experimentally by amputating just one flagellum of the biflagellate alga, and by leaving the second one intact, as  described in Ref. \cite{non-equal1}. 
We note that in this case the regeneration scenario is more complicated, as compared to the case when both flagella are removed (see Ref. \cite{c}). Here, the intact flagellum first shortens linearly in time  while the amputated one regenerates. This way the two flagella attain an equal, intermediate length. Then both grow and  eventually approach
their initial length at the same rate. However,  the time required to reach an equal intermediate length can be quite long, i.e.,   20 to 40 minutes \cite{non-equal1}. It  can become even longer if certain chemicals (e.g., colchicine) are added after a deflagellation, which inhibit the regeneration process \cite{non-equal1,non-equal2}. Therefore, there is a time window in which both flagella have distinctly different lengths. 
Within this time window, the coupled phases
 $\theta_1(t)$ and $\theta_2(t)$ will undergo a stochastic evolution -- each
 at its own temperature  $T^{(1)}_{\rm eff}$ or $T^{(2)}_{\rm eff}$, respectively.
 Such a  system is no longer characterized by a unique effective temperature $T_{\rm eff}$.
One expects that
 the Fokker-Planck equation \cite{kalm} associated with the Langevin equations \eqref{phases}
 will have a non-trivial, current-carrying steady-state solution.
In the following we shall refer to the case of unequal temperatures
 as an out-of-equilibrium case, keeping in mind, of course, that the original physical system is not in equilibrium with its bath, even for $T^{(1)}_{\rm eff} = T^{(2)}_{\rm eff}$.

 Viewed from a different perspective,  which is perhaps equally important
 due to certain other applications (see Ref. \cite{m3} for a discussion)
 we note that  the minimal  model in Eq. \eqref{phases}
 with a unique effective temperature $T_{\rm eff}$ represents the so-called Sakaguchi model \cite{12a} which is
a noisy version of the celebrated Kuramoto model of coupled oscillators
(see, e.g.,  Refs. \cite{11a,m2,11,11b}).
Its generalization to the case of
two different temperatures \cite{25} emerges  naturally within the present context of the synchronization of beating, non-identical flagella.
We note parenthetically that recently the  stochastic evolution of systems with several temperatures
was intensively studied  and a wealth of interesting out-of-equilibrium phenomena has been predicted (see, e.g., Refs. \cite{12,13,14,15,16,17,18,19} and references therein).
To the best of our knowledge, the issue of
synchronization under out-of-equilibrium conditions in general, and in a system with two degrees of freedom exposed to  two different effective temperatures in particular, has  not yet been  addressed.
Inter alia, this motivates our quest for  synchrony in a minimal model with two different effective temperatures.

\begin{figure*}
                \includegraphics[width=0.8\linewidth]{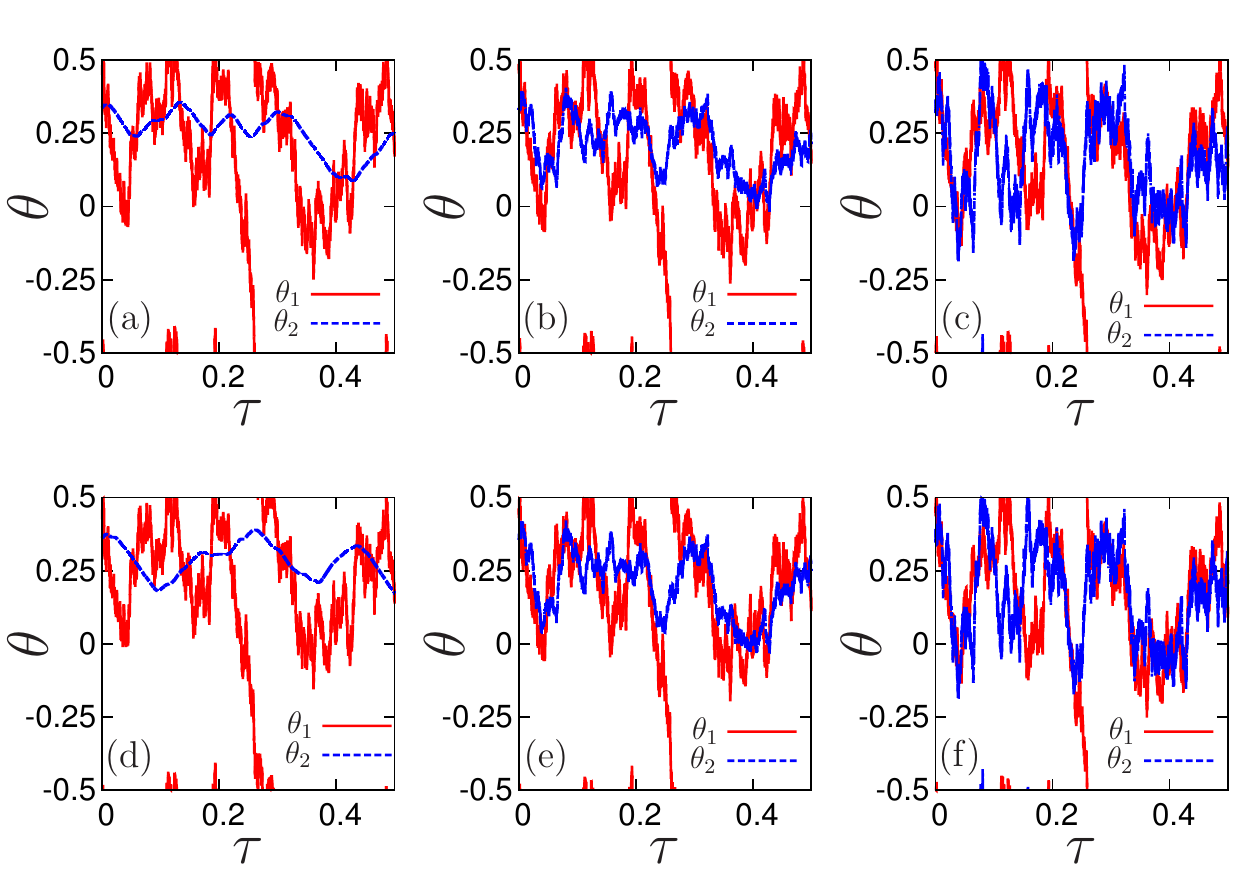}
                \caption{Individual realizations of  (short) trajectories $\theta_1(t)$ and $\theta_2(t)$ (recall that $\theta_1(t),\theta_2(t) \in (-1/2,1/2)$ are periodic quantities), defined in the reference frame rotating with the frequency $\nu$,  for $T_{\rm eff}^{(1)} = 0.5$  {\rm Hz}
as functions of the reduced, dimensionless time $\tau = T_{\rm eff}^{(1)} \, t$. The observation time  is ${\cal S} = 1$ {\rm sec} only.
Panel (a): $\varepsilon = 0.5$ {\rm Hz} and $T_{\rm eff}^{(2)} = 0$, i.e., the rotary motor of the second flagellum is perfect in that it does not fluctuate.
Panel (b): $\varepsilon = 0.5$ {\rm Hz} and $T_{\rm eff}^{(2)} = 0.1$ {\rm Hz}.
Panel (c): $\varepsilon = 0.5$ {\rm Hz} and $T_{\rm eff}^{(2)} = 0.5$ {\rm Hz}.
Panel (d): $\varepsilon = 1$ {\rm Hz} and $T_{\rm eff}^{(2)} = 0$.
Panel (e): $\varepsilon = 1$ {\rm Hz} and $T_{\rm eff}^{(2)} = 0.1$ {\rm Hz}.
Panel (f): $\varepsilon = 1$ {\rm Hz} and $T_{\rm eff}^{(2)} = 0.5$ {\rm Hz}.
The individual trajectories $\theta_1(t)$ and $\theta_2(t)$ exhibit a noisy dynamics,
but nonetheless evolve alongside each other for rather extended periods of time. 
This is precisely the stochastic phase locking phenomenon described  in Ref. \cite{10}
for the case of equal temperatures. It is inferred by following the time evolution of the phase difference $\Delta_t = \theta_1(t) - \theta_2(t)$ in experiments and in numerical simulations.
Here, we observe that this kind of stochastic synchronization \cite{10} degrades if $T_{\rm eff}^{(1)} \neq T_{\rm eff}^{(2)}$. Indeed, the stochastic phase locking
is seemingly strongest in the case of equal temperatures (panels (c) and (f)).
It is less pronounced for the combination $T_{\rm eff}^{(1)} = 0.5$ {\rm Hz} and $T_{\rm eff}^{(2)} = 0.1$ {\rm Hz}
(panels (b) and (e)), and it is weakest for $T_{\rm eff}^{(1)} = 0.5$ {\rm Hz} and $T_{\rm eff}^{(2)} = 0$ (panels (a) and (d)),
for which the periods of synchronization in the dynamics of $\theta_1(t)$ and $\theta_2(t)$ are hardly visible.
               }
                \label{fig:fig2}
\end{figure*}


Here, we focus on the  stochastic evolution of the phases $\theta_1(t)$ and $\theta_2(t)$ of two coupled oscillators, which obeys
the minimal model in Eq. \eqref{phases} with the covariance
 functions of the noise terms of the form
\begin{align}
\label{noises}
\overline{ \zeta_i(t) \zeta_j(t)} = 2 \, \delta_{i,j} \, T^{(i)}_{\rm eff} \, \delta(t-t') \,, \,\,\, i,j = 1,2 \,,
\end{align}
where $T^{(1)}_{\rm eff}$ and $T^{(2)}_{\rm eff}$ are, in  general, not equal.
For simplicity, we assume that the natural frequencies of both oscillators are the same, $\nu_1 = \nu_2 = \nu$.
On one hand, this assumption appears to be justified because
the experimentally observed difference of the natural frequencies is indeed rather small
(see above)  \cite{10}, so that  in a first approximation it can be neglected.
On the other hand, this assumption allows us to disentangle the effects of an out-of-equilibrium 
active noise from the effects caused by a possible, albeit small, difference of the natural  frequencies  $\nu_1$ and $\nu_2$.

We demonstrate, both analytically and numerically,
 that in such a 
 system an emerging steady-state is characterized by a nonzero
  current $j(\theta_1,\theta_2)$ 
  in the frame of reference rotating with  frequency $\nu$ (note that $j(\theta_1,\theta_2) \equiv 0$ 
  when  $T^{(1)}_{\rm eff} =  T^{(2)}_{\rm eff}$). This current, which is the same for both 
phases, with an amplitude depending on the instantaneous values of $\theta_1(t)$ and $\theta_2(t)$, 
  sustains  a synchronized  time evolution of the rates 
  $\dot{\theta}_1(t)$ and $\dot{\theta}_2(t)$ at which the phases change, i.e., it produces correlated drifts of phases.
At the same time, we realize that  the stochastic phase locking seen in Ref. \cite{10}  degrades
 for unequal effective temperatures (see Fig. \ref{fig:fig2}) and is weakest in the case that one of the effective temperatures equals zero,    i.e., that the corresponding rotary motor does not fluctuate. 
 In order to quantify  the degree of  synchronization,  based on the correlated drifts of phases, 
we define a characteristic order parameter $Q$, which 
measures the relative amount of the novel, out-of-equilibrium  synchronization mechanism and
vanishes if the  effective temperatures of the noise terms  become equal.
This definition of $Q$ is
based on the explicit expression derived here for the steady-state current $j(\theta_1,\theta_2)$ and hence, it represents a property averaged over the statistical ensemble of the trajectories of $\theta_1(t)$ and $\theta_2(t)$.  In experiments, however, it is often not possible to garner a sufficiently large statistical sample in order to carry out this kind of averaging. For this reason, 
 we propose an analogous order parameter  $Q_{\cal S}$
defined on the level of a single-realization of  the trajectories $\theta_1(t)$ and $\theta_2(t)$ tracked within a time interval $(0,{\cal S})$.  We show, via numerical 
simulations, that both definitions lead to consistent results,  i.e., $ Q_{\cal S} \to Q$ in the limit of
an  unlimited long observation time ${\cal S}$. This result, inter alia, shows that  the system under study is ergodic, which cannot be expected {\em a priori}, especially if $T^{(1)}_{\rm eff} \neq T^{(2)}_{\rm eff}$.

The outline of the paper is as follows. In Sec. \ref{sec1} we present our main results obtained for the model defined by Eqs. \eqref{phases} and \eqref{noises}. We present here explicit expressions for the probability density function in the steady-state, the steady-state current and an ensemble-averaged order parameter. Further on, we introduce analogous quantities for individual realizations of $\theta_1(t)$ and $\theta_2(t)$. In Sec. \ref{sec2} we discuss the behavior of the ensemble-averaged quantities and of their counterparts defined for a single realization of noises, and outline some perspectives for future research. Details of calculations are relegated to the Appendix \ref{sec3}. Here, we
provide the Fokker-Planck equation associated with the minimal Langevin model in 
Eqs. \eqref{phases} and (\ref{noises}), and present its solution in the limit $t \to \infty$. We also
describe our numerical approach,  which is based on the discretization of the Langevin equations in Eq.~\eqref{phases}.

\section{Results}
\label{sec1}

\subsection{Ensemble-averaged properties.}

{\bf Probability density function in the steady state}. Our main analytical result is an exact expression for the joint probability density function (pdf)
$P(\theta_1,\theta_2)$,  which is the steady-state solution of the Fokker-Planck equation  (see Eq. (\ref{FP}) in the Appendix \ref{sec3}) associated with a system of two coupled Langevin equations (Eq.~\eqref{phases}), and
with the noise terms defined by Eq.  (\ref{noises}):
\begin{align}
\label{dist}
P(\theta_1,\theta_2) = \frac{1}{Z} \exp\left(\frac{\varepsilon}{2 \overline{T}_{\rm eff}} \cos\Big(2 \pi \left(\theta_1 - \theta_2\right)\Big)\right) \,,
\end{align}
where $\overline{T}_{\rm eff}$ is the "mean" effective temperature \cite{35}
$\overline{T}_{\rm eff} = \left(T^{(1)}_{\rm eff} + T^{(2)}_{\rm eff}\right)/2$, while
$Z$ insures that $P(\theta_1,\theta_2)$ is properly normalized, i.e.,
$\int_{-1/2}^{1/2} \int^{1/2}_{-1/2} d\theta_1 d\theta_2 P(\theta_1,\theta_2) = 1$.
This normalization constant can be calculated exactly and is given by
$Z  = {\rm I}_0\left(\varepsilon/(2 \overline{T}_{\rm eff}) \right)$,
where ${\rm I}_0(x)$ is a modified Bessel function of the first kind.  We note that  Eq. \ref{dist} here
represents a particular case of a more general result derived in Ref. \cite{14}.

\begin{figure*}
\begin{center}
\includegraphics[width=0.49\textwidth,height=0.36\textwidth]{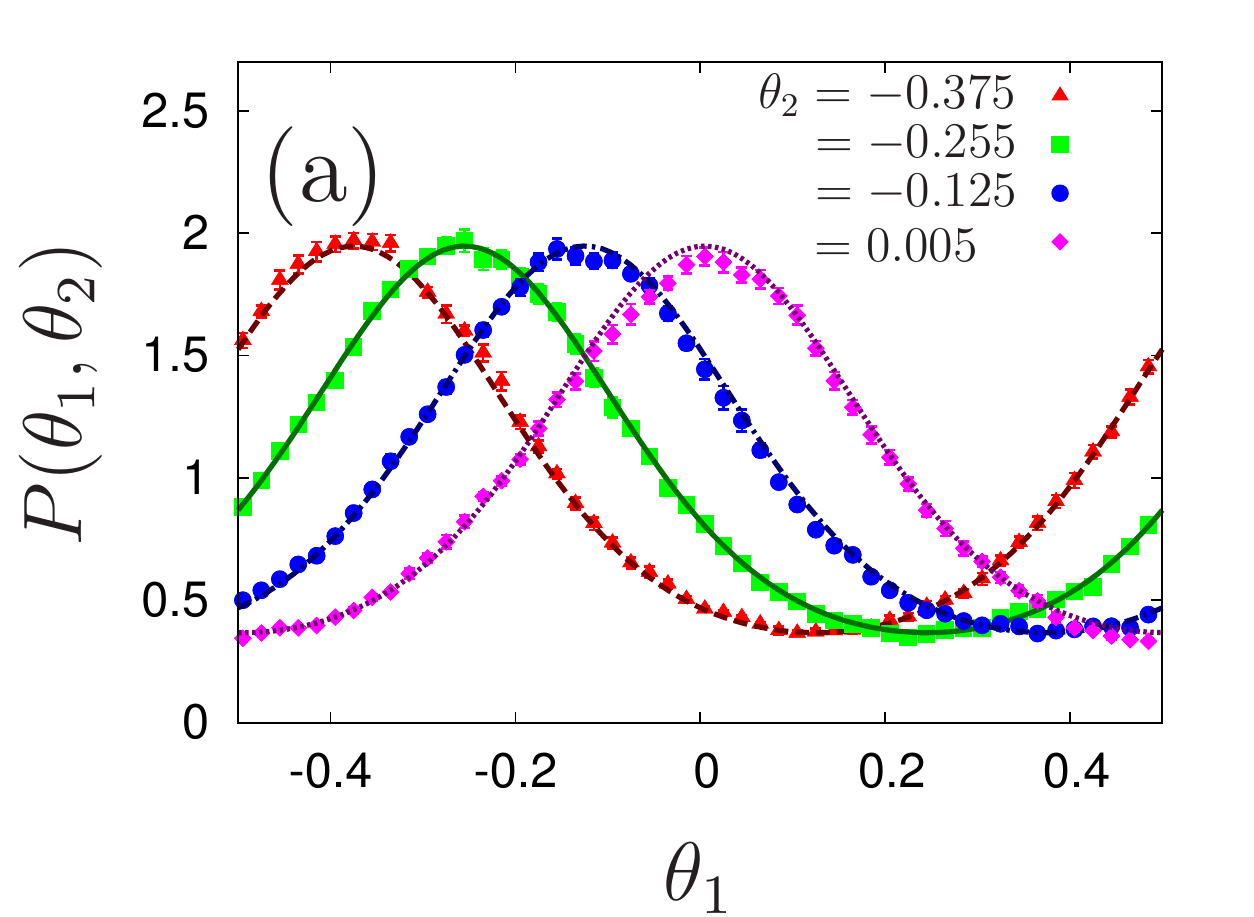}\hspace{-10pt}
\includegraphics[width=0.49\textwidth,height=0.36\textwidth]{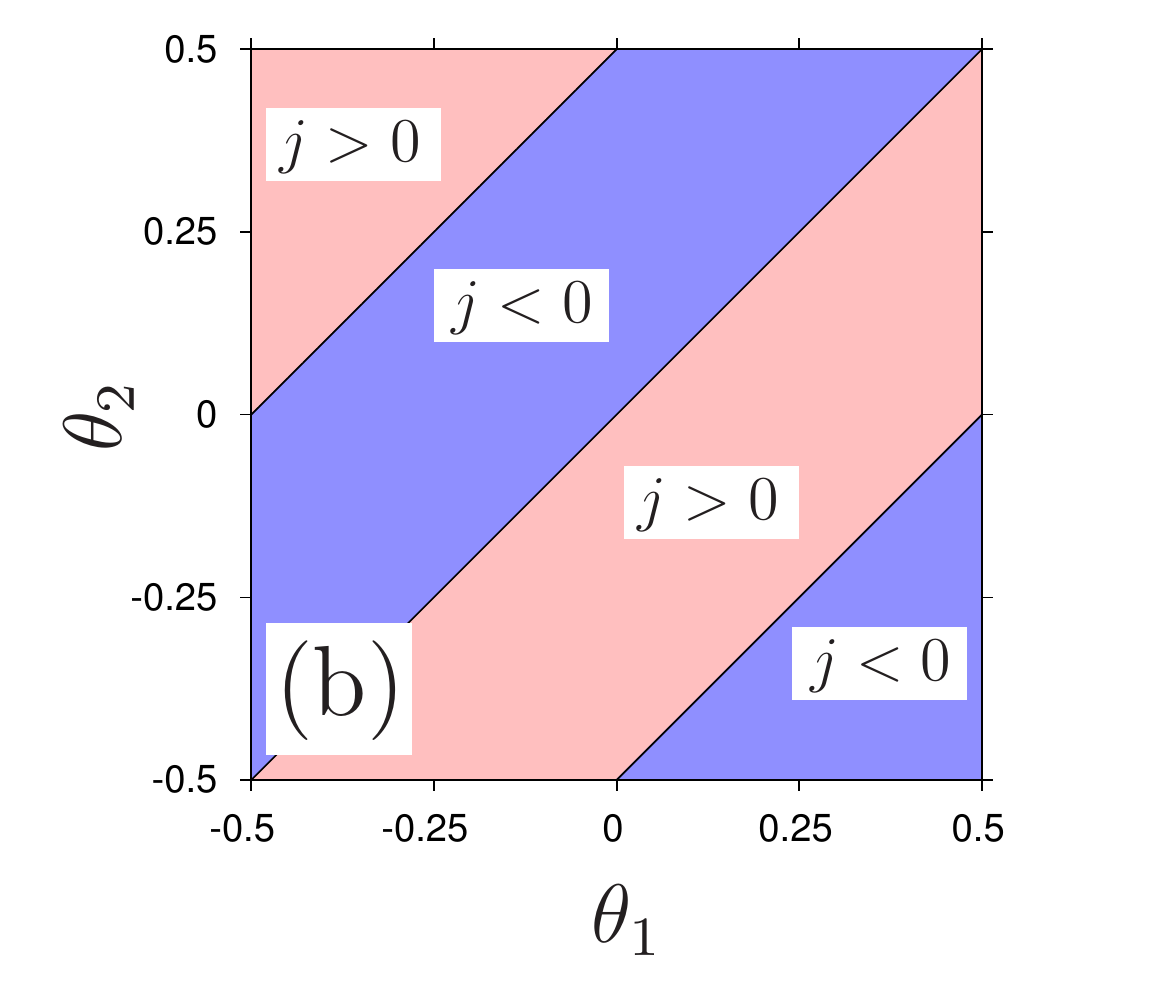}
\includegraphics[width=0.49\textwidth,height=0.36\textwidth]{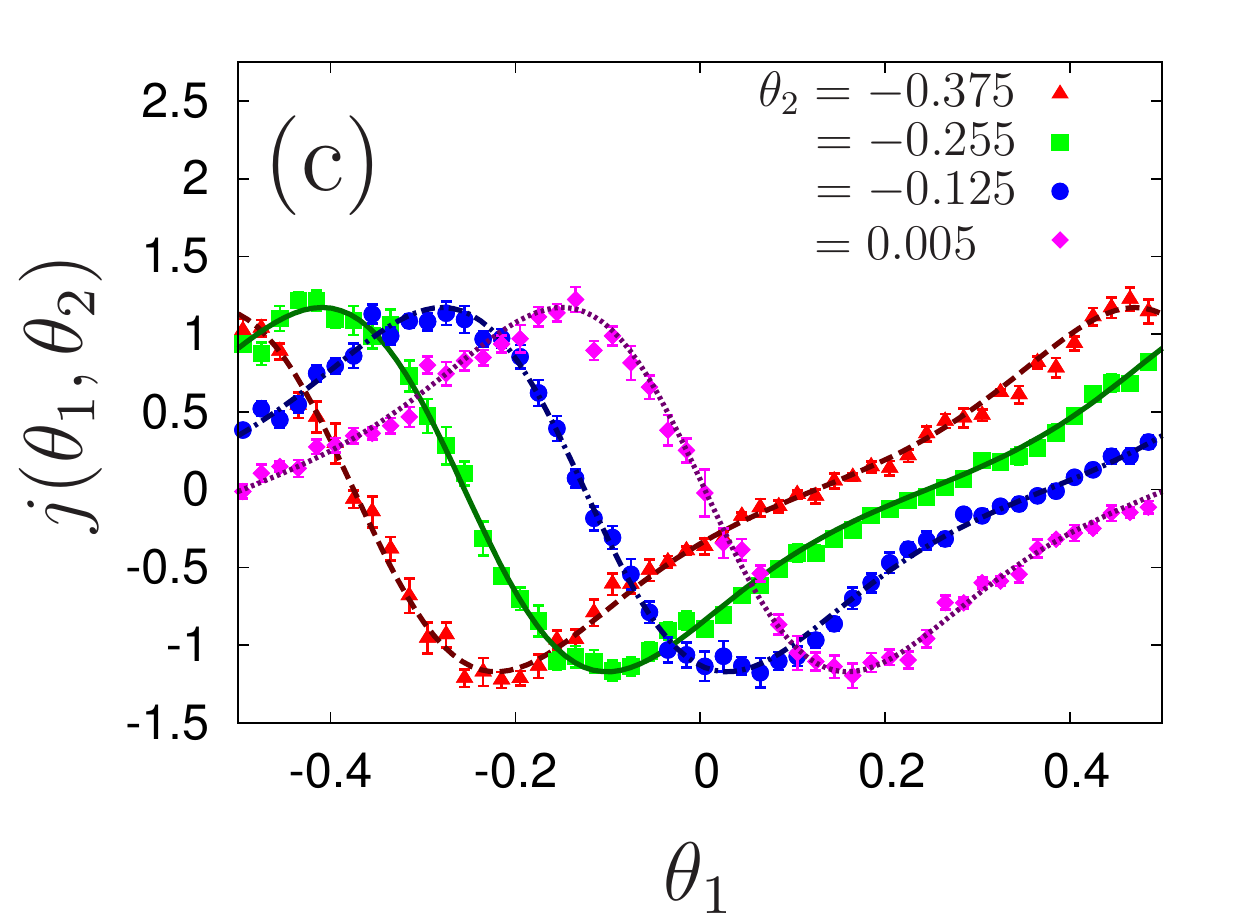}\hspace{-15pt}
\includegraphics[width=0.49\textwidth,height=0.36\textwidth]{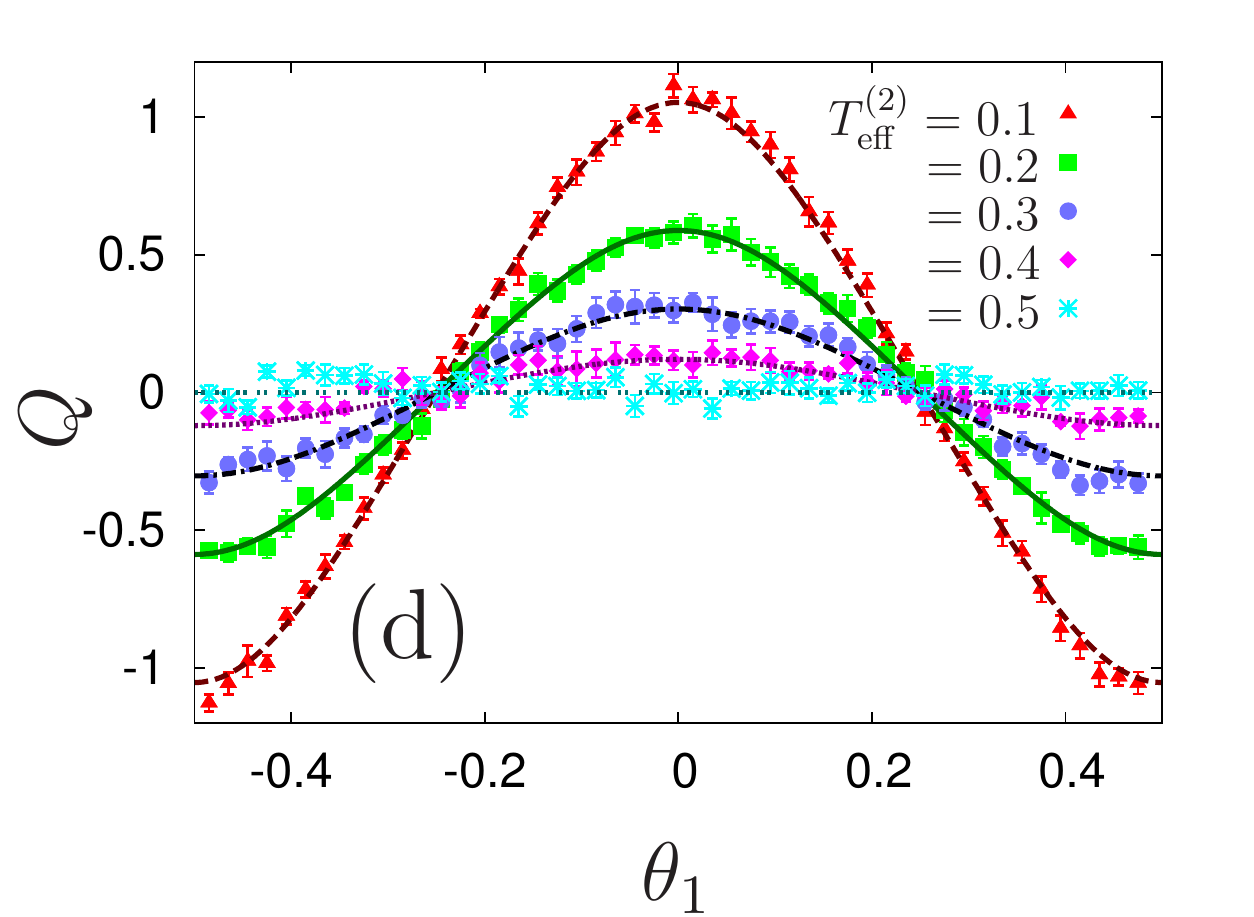}
\caption{
Ensemble-   versus time-averaged properties. 
Panel (a): The pdf $P(\theta_1,\theta_2)$ as a function of $\theta_1$ for four values of $\theta_2$.
The coupling parameter is $\varepsilon = 0.5$  {\rm Hz}, $T_{\rm eff}^{(1)} = 0.5$  {\rm Hz}, and $T_{\rm eff}^{(2)} = 0.1$ {\rm Hz}, consistent with the observations made in Refs. \cite{10,c}.
The curves correspond to the analytical prediction made in Eq. \eqref{dist}.
The symbols represent the results of the numerical simulations (see Sec. \ref{sec2}), 
based on a single-realization, time-averaged $P_{\cal S}(\theta_1,\theta_2)$ (Eq. \eqref{PP}) with ${\cal S} = 10^{4}$ {\rm sec}  (such that for a typical value of the frequency $\nu \simeq 47 \, {\rm Hz}$ \cite{10}, each flagella makes, on average, $4.7 \times 10^{5}$ full beats).
Panel (b): Sign of  the  out-of-equilibrium current
$j(\theta_1,\theta_2)$ (Eq. \eqref{j}), which causes synchronized "drifts" of phases $\theta_1(t)$ and $\theta_2(t)$, on the periodic
$(\theta_1,\theta_2)$-plane for $T_{\rm eff}^{(1)} > T_{\rm eff}^{(2)}$.
The black lines $\theta_1 = \theta_2$ and $|\theta_1 - \theta_2| = 1/2$ correspond to
a change of sign, i.e., $j(\theta_1,\theta_2) = 0$.
Panel (c): The out-of-equilibrium current
$j(\theta_1,\theta_2)$  as a function of $\theta_1$ for four values of 
$\theta_2$. The coupling parameter is $\varepsilon = 0.5$ {\rm Hz},  $T_{\rm eff}^{(1)} = 0.5$ {\rm Hz}, and $T_{\rm eff}^{(2)} = 0.1$ {\rm Hz}.
The curves correspond to the analytical prediction made in Eq. \eqref{j}. The symbols represent the results of numerical simulations  for a single-realization time-averaged current $ j_{\cal S}(\theta_1,\theta_2)$ in Eq. \eqref{jj} with $ {\cal S} = 10^{6}$ {\rm sec}.
Panel (d): Dimensionless order parameter $Q$  as a function of $\theta_1$ for $T_{\rm eff}^{(1)}  = 0.5$ {\rm Hz},
coupling parameter $\varepsilon =0.5$ {\rm Hz}, and five values of $T_{\rm eff}^{(2)}$.
The curves correspond to the analytical prediction made
in Eq. \eqref{Q},  which defines the ensemble-averaged order parameter $Q$. The symbols represent the results of 
the numerical simulations for the time-averaged order parameter $Q_{\cal S}$ (Eq. \eqref{QQ}), based on a single realization of the noises with ${\cal S} = 10^{4}$ {\rm sec} (i.e., approximately $2.8$ hours). Note that for $T_{\rm eff}^{(2)} = 0.5$ {\rm Hz}, i.e., in the case of equal effective temperatures, both the ensemble-averaged order parameter $Q$ and its  time-averaged counterpart $Q_{\cal S}$ are equal to zero.
\label{fig:fig1}
}
\end{center}
\end{figure*}

Since
both natural frequencies $\nu_1$ and $\nu_2$ are taken to be equal, the steady-state 
solution $P(\theta_1,\theta_2)$ depends on the phases only via the phase difference, 
and thus becomes independent of $\nu = \nu_1=\nu_2$ (see Eq. \eqref{dist}). 
At first glance, the latter property appears to be 
 somewhat astonishing, but it 
can be readily understood once one notices that the time evolution of the phase difference in the Langevin equations \eqref{phases} becomes independent of $\nu$ if both natural frequencies are equal to each other. 
This means that $P(\theta_1,\theta_2)$ is defined in the frame of reference rotating with the
unique frequency $\nu$. Naturally,
 the maximum of $P(\theta_1,\theta_2)$
occurs for $\theta_1 = \theta_2$, regardless of the relation between the temperatures  $T^{(1)}_{\rm eff}$ and
 $T^{(2)}_{\rm eff}$. For $\varepsilon/(2 \overline{T}_{\rm eff}) \to \infty$, the pdf turns into 
 a delta function of the difference $\theta_1 - \theta_2$.  Figure \ref{fig:fig1}(a) provides 
 the pdf $P(\theta_1,\theta_2)$ 
  (Eq. \eqref{dist}) as a function of $\theta_1$ for several fixed values of $\theta_2$.

{\bf Out-of-equilibrium current}. A remarkable feature of the minimal model with two different temperatures is,  that 
in the non-equilibrium steady-state a nonzero current {\bf J} occurs.  This is a well-known aspect for 
 stochastic dynamics of coupled components, each evolving at its own temperature (see, e.g., Refs. \cite{12,13,14,15,16,17,18,19}). However, in the case at hand this nonzero current has a peculiar form due to the fact that the coupling term in Eq. (\ref{phases}) is a periodic function of the phase difference.
The components $J_1$ and $J_2$ of this current can be inferred directly from the Fokker-Planck equation \eqref{FP} (see Appendix \ref{sec3}, Eq. (\ref{J2})).
They obey
\begin{align}
\label{J}
J_1 &=  J_2 = j(\theta_1,\theta_2) + \nu P(\theta_1,\theta_2) \; .
\end{align}

The expression on the right-hand-side (rhs) of Eq. \eqref{J} contains the trivial
term $\nu P(\theta_1,\theta_2)$, which is the same for both components as could be expected on general grounds. It 
appears due to the constant drift term $\nu = \nu_1 = \nu_2$ on the rhs of the 
Langevin equations \eqref{phases}. In addition, there is a non-trivial contribution $j(\theta_1,\theta_2)$, 
 which is  a steady-state current 
in the frame rotating with the unique frequency $\nu$; it reads
\begin{align}
\label{j}
j(\theta_1,\theta_2) 
 &= - \pi \varepsilon \frac{\Delta T_{\rm eff}}{2 \overline{T}_{\rm eff}} \sin\left(2 \pi (\theta_2 - \theta_1)\right) P(\theta_1,\theta_2)  \,,
 \end{align}
with $\Delta T_{\rm eff} = T^{(1)}_{\rm eff} - T^{(2)}_{\rm eff}$. Rather unexpectedly,  
$j(\theta_1,\theta_2)$ appears also to be the {\em same} for both components $J_1$  and $J_2$ of the current $\bf J$,  due to the form of the pdf in Eq. \eqref{dist}.

The mean out-of-equilibrium current,
\begin{align}
\label{av}
\langle j(\theta_1,\theta_2) \rangle \equiv \int^{1/2}_{-1/2} \int^{1/2}_{-1/2} d\theta_1 d\theta_2 \, j(\theta_1,\theta_2)  = 0 \,,
\end{align}
vanishes such that, due to $\langle P(\theta_1,\theta_2) \rangle = 1$, 
$\langle J_1\rangle = \langle J_2 \rangle = \nu$. 
One can straighforwardly check that  $\int^{1/2}_{-1/2} d\theta_1 j(\theta_1,\theta_2) \; = \; 0 \; = \; 
\int^{1/2}_{-1/2} d\theta_2 j(\theta_1,\theta_2)$.
On the other hand, $j(\theta_1,\theta_2)$ is not equal to zero locally (except
for $\theta_1 = \theta_2$ and $|\theta_1 - \theta_2| = 1/2$, where the current changes sign),
and its sign and amplitude depend on the precise values of the phases $\theta_1$ and $\theta_2$. In Fig. \ref{fig:fig1}(b), for a particular example with $T_{\rm eff}^{(1)} >  T_{\rm eff}^{(2)}$, we present a "phase chart" for the sign (i.e., the direction) of the out-of-equilibrium current $j(\theta_1,\theta_2)$ in the periodic $(\theta_1,\theta_2)$-plane.
Further on,  in Fig. \ref{fig:fig1}(c) we show the current $j(\theta_1,\theta_2)$ (Eq. \eqref{j}) as a function of $\theta_1$ for several fixed values of $\theta_2$, which also provides insight into its amplitude.


 {\bf Out-of-equilibrium synchronization.} Equations (\ref{J}) and (\ref{j})  demonstrate that  in a minimal model with two different effective temperatures, 
in addition to a stochastic phase locking
of the coupled phases $\theta_1$ and $\theta_2$ (as observed in Ref. \cite{10}), 
there is a different
synchronization mechanism (based on the out-of-equilibrium current $j(\theta_1,\theta_2)$)
which manifests itself via
 drifts of the phases. These drifts are correlated in that they have the same sign (i.e., direction)  
 and the same amplitude for both phases. The actual direction of such drifts depends
on the sign of $\Delta T_{\rm eff}$, as well as on the relative positions of $\theta_1$
and $\theta_2$  with respect to each other. In order to
illustrate this behavior, we suppose  $\Delta T_{\rm eff} > 0$ and
 $0 < \theta_2 - \theta_1 < 1/2$. In this case, according to Eq. (\ref{j}),
both $\theta_1$ and $\theta_2$ experience a drift  in the negative direction up to the time at which,
due to the thermal noise, the phase difference exceeds the value $1/2$ so that  $\theta_2 - \theta_1 > 1/2$.
Then, both $\theta_1$  and $\theta_2$  revert the  direction of their drift.
Once $\theta_1$  and $\theta_2$ interchange their positions, such that $-1/2 < \theta_2 - \theta_1 < 0$,
the current $j(\theta_1,\theta_2)$  changes  sign and turns positive, so
that both $\theta_1$ and $\theta_2$  drift in the positive direction. Once $\theta_2 - \theta_1 < -1/2$,
the drift direction  again changes  sign and becomes negative.

{\bf An order parameter in the steady-state.} In order to quantify this novel synchronization mechanism, 
and also in order to render it observable either experimentally or in numerical simulations,
one has  to introduce a meaningful order parameter.
In view of the above discussion, for a minimal model
with two effective temperatures the latter should be associated with the steady-state current $j(\theta_1,\theta_2)$.
As in general, there is, however, some liberty in choosing this parameter.
Here  we define an order parameter by integrating the out-of-equilibrium current
$j(\theta_1,\theta_2)$ over  $\theta_2$ across  half of the domain in which this variable is defined, and dividing the result by the mean effective temperature. This gives the following \textit{dimensionless} order parameter
(see Eqs. \eqref{dist} and \eqref{j})
\begin{align}
\label{Q}
Q(\theta_1) &= \frac{1}{\overline{T}_{\rm eff} }   \int^{1/2}_{0} d\theta_2  \, j(\theta_1,\theta_2) \nonumber\\
&=  \frac{\Delta T_{\rm eff}}{\overline{T}_{\rm eff}} \frac{{\rm sinh}\left((\varepsilon/(2 \overline{T}_{\rm eff}) \cos\left(2 \pi \theta_1\right) \right)}{{\rm I}_0\left(\varepsilon/(2 \overline{T}_{\rm eff}) \right)} \,.
\end{align}
This order parameter $Q$  is a function of the phase $\theta_1$ and
depends on the coupling parameter $\varepsilon$ as well as  on the values of the effective temperatures. It vanishes 
 for $\varepsilon \to 0$, for $\overline{T}_{\rm eff} \to \infty$, and for $T_{\rm eff}^{(1)} \to T_{\rm eff}^{(2)}$,  the latter limit being characteristic of the transition to the equilibrium setup.  
 The order parameter also vanishes for $\theta_1 = \pm \pi/4$. 

It is rewarding to determine  the asymptotic behavior of $Q$ in several particular limits.
For instance, in the high-temperature limit one has
\begin{align}
Q(\theta_1) \; \simeq \;  \frac{\Delta T_{\rm eff}}{2 \overline{T}^2_{\rm eff}} \varepsilon \cos\left(2 \pi \theta_1\right) \,, \,\,\,
\overline{T}_{\rm eff}  \gg \varepsilon/2,
\end{align}
which reduces to
\begin{align}
Q(\theta_1) \simeq \frac{2 \,  {\rm sign}\left(T_{\rm eff}^{(1)} - T_{\rm eff}^{(2)}\right)}{\max\left(T_{\rm eff}^{(1)},T_{\rm eff}^{(2)}\right)} \, \varepsilon \cos\left(2 \pi \theta_1\right) \,,
\end{align}
if one of the effective temperatures is much higher than the other one.

In the  opposite limit   $\overline{T}_{\rm eff}  \ll (\varepsilon \cos(2 \pi \theta_1))/2$,
which can be reached either via a sufficiently strong coupling $\varepsilon$
(and for such values of $\theta_1$ for which $\cos(2 \pi \theta_1)$ is nonzero) or
if both temperatures are sufficiently small.
In these cases the system is close to the realm of the standard noiseless Kuramoto model and one finds 
directly from Eq. \eqref{Q} 
that, in  leading order in the parameter $\varepsilon/(2 \overline{T}_{\rm eff}) \to \infty$,  the order parameter $Q$  varies as
\begin{align}
Q(\theta_1) \; \simeq \; \frac{\Delta T_{\rm eff}}{2 \overline{T}^{3/2}_{\rm eff}} (\pi \varepsilon)^{1/2}
\exp\left(- \frac{\varepsilon}{\overline{T}_{\rm eff} } \sin^2\left(\pi \theta_1\right)\right).
\end{align}
In these limits $T_{\rm eff}^{(1,2)} \to 0$ and for any $\theta_1\neq 0$, $Q(\theta_1)$ 
is exponentially small.
For $\theta_1 = 0$,  the exponential factor in the latter expression equals $1$
and hence the order parameter $Q$ varies algebraically as function of
the effective temperatures and the coupling parameter $\varepsilon$: 
\begin{equation}
Q(\theta_1=0) \simeq \frac{\Delta T_{\rm eff} (\pi \varepsilon)^{1/2}}{2 \overline{T}^{3/2}_{\rm eff}} \,,
\end{equation}
i.e.,  it is a non-analytic function of the coupling parameter and the effective temperatures.

\subsection{Time-averaged properties for individual realizations of noises.}

 It is not always possible to generate a statistical sample of 
large enough size, either in experiments 
or in numerical simulations, which allows one to average over an ensemble of trajectories.
To this end, we present alternative definitions for the pdf $P(\theta_1,\theta_2)$, the 
current $j(\theta_1,\theta_1)$, and the order parameter $Q$, based on their time-averaged counterparts.\\

 {\bf The pdf for a single realization of noises}. The pdf $P(\theta_1,\theta_2)$ (Eq. (\ref{dist}))  obeys $P(\theta_1,\theta_2) = \lim_{{\cal S} \to \infty} P_{\cal S}(\theta_1,\theta_2)$ with
\begin{align}
\label{PP}
P_{\cal S}(\theta_1,\theta_2) = \frac{1}{\cal S} \int^{\cal S}_0 dt \, \delta\left(\theta_1(t) - \theta_1\right) \delta\left(\theta_2(t) - \theta_2\right) \,,
\end{align}
 where $\theta_1(t)$ and $\theta_2(t)$ are two individual realizations of the trajectories of the phases, 
corresponding to the solutions of the Langevin equations (Eq. \eqref{phases}) for
a given realization of the noises $\zeta_1(t)$ and $\zeta_2(t)$ in Eq. \eqref{noises}. 
In Eq. \eqref{PP}, $P_{\cal S}(\theta_1,\theta_2)$ is the total number of simultaneous occurrences, within the time interval $(0,{\cal S})$, of two given realizations of the trajectories $\theta_1(t)$ and $\theta_2(t)$ at the positions  $\theta_1$ and $\theta_2$, respectively, divided by the observation time ${\cal S}$. 
If the system under study is ergodic, as it is the case (see Sec. \ref{sec2}), the ensemble average and the time average 
yield identical results, such that, in the limit ${\cal S} \to \infty$, 
$ P_{\cal S}(\theta_1,\theta_2)$ should attain $P(\theta_1,\theta_2)$. 

{\bf Time-averaged current.} We introduce the current $j_{\cal S}(\theta_1,\theta_2)$ as an average
over the observation time $ {\cal S}$:
\begin{align}
\label{jj}
 j_{\cal S}(\theta_1,\theta_2) &= -   \frac{1}{\cal S} \int^{\cal S}_0 dt \, \dot{\theta}_1(t) \delta\left(\theta_1(t) - \theta_1\right) \delta\left(\theta_2(t) - \theta_2\right) \nonumber\\
& = -  \frac{1}{\cal S} \int^{\cal S}_0 dt \, \dot{\theta}_2(t) \delta\left(\theta_1(t) - \theta_1\right) \delta\left(\theta_2(t) - \theta_2\right) \,,
\end{align}
where $\theta_1(t)$ and $\theta_2(t)$ obey Eq. \eqref{phases} with $\nu$ set  to zero. Note that the expressions in the first and the second line in Eq. \eqref{jj} correspond to the components of the current {\bf J} and 
differ with respect to the time derivative of the phases, i.e., $\dot{\theta}_1(t)$ or $\dot{\theta}_2(t)$. As we have shown above, the components of the ensemble-averaged current are exactly equal to each other. We thus expect (and verify via numerical simulations) that the same holds for their introduced time-averaged counterparts.

{\bf Time-averaged order parameter $Q_{\cal S}$}.  We integrate the expression in 
the first line on the rhs of Eq. \eqref{jj} over  $\theta_2$ 
across one half of the domain in which this variable is defined. Dividing the result by 
the mean effective temperature (see the definition of $Q$ in Eq. \eqref{Q}), we obtain
\begin{align}
\label{QQ}
Q_{\cal S} &= -   \frac{1}{{\cal S} \overline{T}_{\rm eff}} \int^{\cal S}_0 dt \, \dot{\theta}_1(t) \delta\left(\theta_1(t) - \theta_1\right)  \nonumber\\
&  \times \int_0^{1/2} d\theta_2 \, \delta\left(\theta_2(t) - \theta_2\right) \nonumber\\
&= -   \frac{1}{{\cal S} \overline{T}_{\rm eff}} \int^{\cal S}_0 dt \, \dot{\theta}_1(t) \delta\left(\theta_1(t) - \theta_1\right) \theta_H\left(\theta_2(t)\right) \,,
\end{align}
where $\theta_H(z)$ is the Heaviside theta-function, which is zero for $z < 0$
and $1$ for $z > 0$. We expect that, similarly to the time-averaged quantity $P_{\cal S}(\theta_1,\theta_2)$ (Eq. \eqref{PP})  and to the time-averaged current in Eq. \eqref{jj}, 
for large observation times $ {\cal S} \to \infty$,
$ Q_{\cal S}$ converges to $Q$ given 
in Eq. \eqref{Q}.

\section{Discussion}
\label{sec2}

In Fig. \ref{fig:fig2} we present appropriately discretized, 
individual realizations of the trajectories $\theta_1(t)$ and $\theta_2(t)$ which consist of $n=10^6$ steps with a 
discrete time step $\delta t = 10^{-6}$ {\rm sec}  (see the Appendix \ref{sec3} for more details)
for two distinct values of the coupling parameter
($\varepsilon = 0.5 \, {\rm Hz}$  for the upper row and $\varepsilon = 1 \,  {\rm Hz}$ for the lower row), for the
fixed effective temperature $T_{\rm eff}^{(1)} = 0.5 \, {\rm Hz}$ of flagellum $1$, and
for  three  temperatures of  flagellum
$2$ ($T_{\rm eff}^{(2)} = 0, 0.1 \,  {\rm Hz}$, and $0.5 \,  {\rm Hz}$).
The case  $T_{\rm eff}^{(2)} = 0.5 \, {\rm Hz}$, i.e., $T_{\rm eff}^{(1)} = T_{\rm eff}^{(2)}$
 corresponds to   the original one
considered in Ref. \cite{10}. In contrast,
the trajectories in panels (a) and (d) of Fig. \ref{fig:fig2} correspond, within the present choices,
to the extreme case of maximal disparity between the temperatures. 
In (a), $\theta_{2}(t)$ corresponds to zero temperature 
(i.e.,  a perfect rotary motor operating with no noise)
and is entrained in random motion by $\theta_{1}(t)$,
which is subject to  random noise. Interestingly, the stochastic phase locking described in Ref. \cite{10} is seemingly
strongest in the case of equal temperatures (panels (c) and (f)),
is less pronounced for the combination  $T_{\rm eff}^{(1)} = 0.5 \,  {\rm Hz}$ and $T_{\rm eff}^{(2)} = 0.1 \, {\rm Hz}$
(panels (b) and (e)) and it is weakest for $T_{\rm eff}^{(1)} = 0.5 \, {\rm Hz}$ and $T_{\rm eff}^{(2)} = 0$ (panels (a) and (d))
for which the periods of synchrony are hardly visible. Therefore, the synchronization observed
in Ref. \cite{10} degrades if the effective temperatures become unequal.

The pdf $P(\theta_1,\theta_2)$ as a function of $\theta_1$ for several values of $\theta_2$ is illustrated in Fig. \ref{fig:fig1}(a)
together with the results of numerical simulations for the time-averaged, single-trajectory quantity  $P_{\cal S}(\theta_1,\theta_2)$ in Eq. (\ref{PP}).
The  very nice agreement between $P(\theta_1,\theta_2)$ and $ P_{\cal S}(\theta_1,\theta_2)$ 
shows indirectly that the system is indeed ergodic. Such an agreement is, however, achieved  
 for trajectories which are
substantially longer than the ones shown in Fig. \ref{fig:fig2}. Here we have used  the same $\delta t = 10^{-6}$ {\rm sec} but
a  larger value   $n = 10^{10}$, so that the observation time  is   ${\cal S} = 10^4$ {\rm sec}. 

We use next
the trajectories provided in Fig. \ref{fig:fig2} in order to obtain the introduced  time-averaged current $ j_{\cal S}(\theta_1,\theta_2)$ (Eq.  \eqref{jj}) for
  individual realizations of $\theta_1(t)$ and $\theta_2(t)$. The results (see the Appendix \ref{sec3} for more details) are presented 
in Fig. \ref{fig:fig1}(c) together with
the ensemble-average of  $j(\theta_1,\theta_2)$ (Eq. \eqref{j}) as obtained from the solution of the Fokker-Planck equation. 
The agreement between the two results is very satisfactory  for $n = 10^{12}$  and $\delta t = 10^{-6}$ {\rm sec}, such that the observation time ${\cal S} = 10^6$ {\rm sec}. For smaller $n$ the data appear more noisy and no conclusive statement on the convergence of $j_{\cal S}(\theta_1,\theta_2)$ to $j(\theta_1,\theta_2)$ can be made.

In Fig. \ref{fig:fig1}(d)
we show $Q$ obtained from Eq. \eqref{Q} 
together with $Q_{\cal S}$ following from Eq. \eqref{QQ}. 
The latter is obtained 
from the trajectories depicted in Fig. \ref{fig:fig2}  (with $\delta t = 10^{-6}$ and $n=10^{10}$, such that ${\cal S}=10^4$ {\rm sec}), as function of 
$\theta_1$ for $\varepsilon = 0.5 \,  {\rm Hz}$, $T_{\rm eff}^{(1)} = 0.5 \,  {\rm Hz}$,   
and three  values of $T_{\rm eff}^{(2)}$. We observe full agreement between 
our theoretical prediction in Eq. \eqref{Q}, which is defined for an ensemble of trajectories, 
and $Q_{\cal S}$ as introduced in Eq. \eqref{QQ}, 
which is defined for a single realization of noises.
This implies that the latter can be conveniently used for a single-trajectory  
analysis of corresponding experimental and numerical data. 
Finally,  we note  that a rather long observation time ${\cal S}=10^4$ {\rm sec} has been used in Fig. \ref{fig:fig1}(d) 
 in order
to demonstrate convergence of the time-averaged order parameter to the ensemble-averaged one. The observation that the order parameter $Q_{\cal S}$ deviates from zero in out-of-equilibrium conditions can be made already 
for more moderate values of $n$, although the data will look more noisy.

In summary,  we have presented a generalization of a minimal model introduced 
in Ref. \cite{10} to  the case in which the phases
in Eq. \eqref{phases} are subject to noises with different amplitudes.
This can be thought of as a noisy Kuramoto  (or Sakaguchi) model of 
two coupled oscillators with distinct effective temperatures. From a physical point of view,
the original model in Ref. \cite{10} has been introduced in order to describe the noisy synchronization 
of two identical flagella of a biflagellate alga. 
Our generalized model is expected to be appropriate for the description of a noisy
synchronization of two flagella having different lengths. Indeed, the analysis in Ref. \cite{c}
has revealed that the noise amplitudes  depend on the length
of the flagella. Viewed from a different perspective, our study provides an, apparently first, 
solvable example for the \text{synchronization} of coupled oscillators under out-of-equilibrium conditions.  
Hence, it opens new perspectives
for a similar analysis of more complicated models, such as a FitzHugh-Nagumo model (see, e.g., Ref. \cite{m3}).
Note that in the  example studied here the difference between the effective temperatures 
is not artificially imposed but emerges naturally.

We have shown, both analytically and numerically,
 that in such a system a very peculiar form of a synchronization of two coupled oscillators takes place.
It is mediated by an emerging, current-carrying steady-state.
More specifically, we have shown that, on top of the synchronization 
of the phases as observed in Ref. \cite{10},  i.e., a stochastic phase locking,
an additional synchronized rotation (drifts) of the phases takes place.
In order to quantify the degree of such a synchronization, we have introduced  a characteristic
order parameter, which vanishes if the effective temperatures 
become equal to each other. This order parameter has
been determined as the average over an ensemble of realizations of
the stochastic evolution of phases, as well as on the  level of an individual realization.
The latter makes
the order parameter suitable for  experimental and numerical analyses, for which 
a sufficiently large statistical sample cannot be formed.
Via numerical simulations we have shown
that both definitions become equivalent in the limit of sufficiently long observation times, 
which also demonstrates the ergodicity of the system under study. Finally, we remark that we expect 
a much richer behavior for the relevant situation in which, in addition to unequal effective temperatures, the natural frequencies are also different. This is a challenging subject for future research.

\appendix
\section{Details of calculations.}
\label{sec3}

{\bf Analytical approach}. We provide the Fokker-Planck equation for the joint pdf
$P(\theta_1,\theta_2)$, associated with the system of two coupled Langevin equations (Eq.~\eqref{phases}) 
with the noise terms defined by Eq.  (\ref{noises}).
The associated Fokker-Planck equation is derived by standard means \cite{kalm} and reads
\begin{align}
\label{FP}
&\dot{P}(\theta_1,\theta_2) =  - {\rm div}{\bf J} =
 - \frac{\partial}{\partial \theta_{1}} \, J_{1} \; - \; \frac{\partial}{\partial \theta_{2}} \, J_{2}
\nonumber\\
&= \Bigg(\frac{\partial}{\partial \theta_1} \Big[T^{(1)}_{\rm eff} \frac{\partial}{\partial \theta_1} + \Big(\pi \varepsilon \sin\left(2 \pi (\theta_1 - \theta_2)\right)  - \nu\Big)\Big] \nonumber\\
& + \frac{\partial}{\partial \theta_2} \Big[T^{(2)}_{\rm eff} \frac{\partial}{\partial \theta_2} - \Big(\pi \varepsilon \sin\left(2 \pi (\theta_1 - \theta_2)\right)  + \nu\Big)\Big] \Bigg) P(\theta_1,\theta_2) \,,
\end{align}
where ${\bf J} = (J_1, J_2)$ is the probability current.
The steady-state solution of Eq. (\ref{FP}) can be determined analytically and is given by Eq. (\ref{dist}) in Sec. \ref{sec1}.

From Eq. (\ref{FP}) we infer the following expressions for the components $J_1$ and $J_2$ of the out-of-equilibrium current ${\bf J}$:
\begin{align}
\label{J2}
J_1 &=  -\Bigg(T^{(1)}_{\rm eff} \frac{\partial }{\partial \theta_1} + \Big(\pi \varepsilon \sin\left(2 \pi \left(\theta_1 - \theta_2\right)\Big) - \nu \right) \Bigg)P(\theta_1,\theta_2) \nonumber\\
J_2 &=  - \Bigg(T^{(2)}_{\rm eff}  \frac{\partial }{\partial \theta_2} - \Big(\pi \varepsilon \sin\left(2 \pi \left(\theta_1 - \theta_2\right)\Big) + \nu \right) \Bigg) P(\theta_1,\theta_2) . \nonumber\\
\end{align}
Inserting the explicit expression of the pdf in Eq. (\ref{dist}) into Eq. (\ref{J2}), and performing differentiations, we obtain Eq. (\ref{J}).

Rewriting the components $J_1$ and $J_2$ of the out-of-equilibrium current $\bf J$ in that frame of reference 
which rotates 
with the frequency $\nu$, for $j(\theta_1,\theta_2)$ (see the definition in Eq. (\ref{j})) we find 
\begin{align}
\label{j2}
j(\theta_1,\theta_2) &=  -\Bigg(T^{(1)}_{\rm eff}  \frac{\partial}{\partial \theta_1} + \pi \varepsilon \sin\left(2 \pi \left(\theta_1 - \theta_2\right)\right) \Bigg) P(\theta_1,\theta_2) \nonumber\\
&= - \Bigg(T^{(2)}_{\rm eff}  \frac{\partial }{\partial \theta_2} - \pi \varepsilon \sin\left(2 \pi \left(\theta_1 - \theta_2\right)\right)\Bigg) P(\theta_1,\theta_2)  \,.\nonumber\\
 &= - \pi \varepsilon \frac{\Delta T_{\rm eff}}{2 \overline{T}_{\rm eff}} \sin\left(2 \pi (\theta_2 - \theta_1)\right) P(\theta_1,\theta_2)  \,.
 \end{align}
 The expression in the last line in Eq. (\ref{j2}) corresponds to our Eq. (\ref{j}).

{\bf Numerical approach.} Here, we provide a brief description of our numerical algorithm.
To this end we rewrite the Langevin equations (Eq. \eqref{phases}) 
in the frame of reference rotating with the frequency $\nu$, 
i.e., we change variables according to $\theta_{1,2} = \tilde{\theta}_{1,2}(t) + \nu t$,
and then we
discretize the time variable $t = n \delta t$, where $n$ is an integer; $\delta t$ is the time-interval 
between the consecutive steps. Without loosing generality, in what follows we set $\nu = 0$ 
and, in order to avoid a clumsy notation, we drop the tilde mark.  Lastly, we recall the standard scaling properties of Gaussian delta-correlated noises $\zeta_{1,2}(t)$ in order to cast the noise terms into a different (but equivalent) form, in which the effective temperatures appear explicitly as amplitudes of the noises \cite{kalm}.
This turns  Eq. \eqref{phases} into recurrence relations of the form
\begin{align}
\label{phases2}
\theta_1(t+ \delta t) &= \theta_1(t) - \delta t \left(\pi \varepsilon \sin\left[2 \pi \Delta_t\right] +  \sqrt{\frac{2  T_{\rm eff}^{(1)}}{\delta t}} \eta_{1}(t)\right) \,, \nonumber\\
\theta_{2}(t + \delta t) &= \theta_2(t) + \delta t \left(\pi \varepsilon \sin\left[2 \pi \Delta_t\right] +   \sqrt{\frac{2  T_{\rm eff}^{(2)}}{\delta t}} \eta_{2}(t)\right) \,,
\end{align}
where $\Delta_t = \theta_1(t) - \theta_2(t)$ is an instantaneous phase difference.
The above recursions  (with dimensionless prefactors of the sine and of the noise terms) allow us to define the values of $\theta_1(t + \delta t)$ and $\theta_2(t + \delta t)$ 
through the values of $\Delta_t$ and the values of the noise terms $\eta_{1,2}(t)$ at the previous moment of time. 
This permits us to sequentially generate individual realizations
of the phases $\theta_1(t)$ and $\theta_2(t)$ of arbitrary duration $t$. The noises $\eta_1(t)$ and $\eta_2(t)$ 
in Eq. \eqref{phases2} are  dimensionless Gaussian random variables, uncorrelated for distinct values of $n$,  with zero mean and  
variances $\sigma^2_{1,2} \equiv 1$, such that the probability density function is given explicitly by $P(\eta_{1,2}) = \exp(-\eta_{1,2}^2/2)/\sqrt{2 \pi}$.
This choice ensures that
for $\varepsilon = 0$  the phases $\theta_1(t)$ and $\theta_2(t)$ undergo
standard diffusive motion on the unit circle with diffusion coefficients $T^{(1)}_{\rm eff}$ and $T^{(2)}_{\rm eff}$, respectively.

Adopting $\delta t = 10^{-6} \, \rm sec$  (which is sufficiently short such that for a typical beating 
frequency of $\nu \simeq 47 \, {\rm Hz}$ (see Ref. \cite{10}) each flagella makes a full beat within roughly  $2.13 \times 10^4$ intervals $\delta t$), we generate two individual realizations of noises,  thereby
building up individual realizations of the trajectories $\theta_1(t)$ and $\theta_2(t)$ which consist of $n=10^6$ steps. As a consequence, within the full observation time  (here ${\cal S} = 1$ {\rm sec}) each flagella makes, on average,  $47$ full beats.
These trajectories are depicted in Fig. \ref{fig:fig2}. 

In order to determine numerically the introduced  time-averaged current $ j_{\cal S}(\theta_1,\theta_2)$ (Eq.  \eqref{jj}) for
  individual realizations of $\theta_1(t)$ and $\theta_2(t)$, we  first appropriately discretize the expression on the rhs of Eq. \eqref{jj} 
and replace the time-derivative $\dot{\theta}_1(t)$ (or $\dot{\theta}_2(t)$) by the finite difference given by  Eq. \eqref{phases2}. 
Then,  we use
the trajectories provided in Fig. \ref{fig:fig2}. The results of such a procedure are presented 
in Fig. \ref{fig:fig1}(c) together with
the ensemble-average of  $j(\theta_1,\theta_2)$ (Eq. \eqref{j}) as obtained from the solution of the Fokker-Planck equation. The order parameter
 $Q_{\cal S}$, as introduced in Eq. \eqref{QQ}, is determined numerically in a similar way.


\section*{Acknowledgments}

The authors acknowledge valuable discussions with M. Dykman, F. Ritort, M. Rosenblum, and S. Ruffo.
The work by A.M. has been  partially  supported by the Polish National Science Center
(Harmonia Grant No. 2015/18/M/ST3/00403).


\begin{thebibliography}{99}
\bibitem{remark} Note that the definition of the covariance of the noise terms used
in Ref. \cite{10} does not contain the usual prefactor $2$ of the temperature  (see, e.g., Ref. \cite{kalm}). Thus,
in our settings $T_{\rm eff}$ is  half the analogous  quantity in Ref. \cite{10}.
The values of $T_{\rm eff}$, which we present in the main text, take this property into account.
\bibitem{22} We note that the amplitude
of the noise of a single flagellum as a function of its length $l$ might be different from that
inferred from the experimental data in Ref.  \cite{c} for a pair of flagella.
\bibitem{25} The dynamics of two coupled Kuramoto oscillators
exposed to two unequal noises
has been already studied (see Refs. \cite{m2,m3,m1}).
However, for this situation the only (albeit quite complicated)
issue, which was considered, concerns the form of the
diffusion coefficient of the phase difference $\Delta_t$.
\bibitem{35} 
The effective temperatures enter the steady-state pdf in Eq. (\ref{dist}) only 
in an additive way, such that there is  no singular behavior in the case that
 they are unequal;  here the only pertinent quantity
 is the mean effective temperature $\overline{T}_{\rm eff}$. We note that this seems to be a general rule (see Eq. (3) in Ref. \cite{14})
 for any form of the coupling term between the phases $\theta_1(t)$ and $\theta_2(t)$ in Eq. \eqref{phases},  
 which depends on the phase difference only.
 As a consequence,  the  standard complex
 order parameter 
 $r \exp(i \Psi) \equiv (\exp(2\pi i \theta_1) + \exp(2\pi  i \theta_2))/2$, where $r$ defines the phase coherence and $\Psi$ the average phase
  \cite{11a,11,11b}, averaged over the pdf given in Eq. \eqref{dist}, is exactly equal to zero.
\end{thebibliography}
\end{document}